\def\year{2021}\relax
\documentclass[letterpaper]{article} 
\usepackage{aaai21}  
\usepackage{times}  
\usepackage{helvet} 
\usepackage{courier}  
\usepackage[hyphens]{url}  
\usepackage{graphicx} 
\usepackage{natbib}  
\usepackage{caption} 
\usepackage{amsmath}
\usepackage{multirow}
\title{Information Retention in the Multi-platform Sharing of Science}

\author {
    Sohyeon Hwang\textsuperscript{\rm 1},
    Emőke-Ágnes Horvát\textsuperscript{\rm 1},
    Daniel M. Romero\textsuperscript{\rm 2} \\
}
\affiliations {
    \textsuperscript{\rm 1}Northwestern University
    \textsuperscript{\rm 2}University of Michigan \\
    sohyeonhwang@u.northwestern.edu, a-horvat@northwestern.edu, drom@umich.edu
}
\begin{document}

\maketitle

\begin{abstract}
The public interest in accurate scientific communication, underscored by recent public health crises, highlights how content often loses critical pieces of information as it spreads online. However, multi-platform analyses of this phenomenon remain limited due to challenges in data collection. Collecting mentions of research tracked by Altmetric LLC, we examine information retention in the over 4 million online posts referencing 9,765 of the most-mentioned scientific articles across blog sites, Facebook, news sites, Twitter, and Wikipedia. To do so, we present a burst-based framework for examining online discussions about science over time and across different platforms. To measure information retention we develop a keyword-based computational measure comparing an online post to the scientific article's abstract. We evaluate our measure using ground truth data labeled by within field experts. We highlight three main findings: first, we find a strong tendency towards low levels of information retention, following a distinct trajectory of loss except when bursts of attention begin in social media. Second, platforms show significant differences in information retention. Third, sequences involving more platforms tend to be associated with higher information retention. These findings highlight a strong tendency towards information loss over time---posing a critical concern for researchers, policymakers, and citizens alike---but suggest that multi-platform discussions may improve information retention overall.
\end{abstract}

\section{Introduction}
Online platforms are increasingly the predominant medium for communicating science to the public \citep{suScienceNewsConsumption2015}. In 2020, the National Science Board noted that the internet had become the most widely used source for science news, with 57\% of Americans citing online platforms as their primary source of science and technology information 
\citep{nationalscienceboardScienceTechnologyPublic2020}. Given this, the accurate communication of scientific research online is a fundamental concern for policy-making, public health, and establishing the legitimacy of scientific work \cite{jamiesonSignalingTrustworthinessScience2019}.

Prior work notes that information often diverges from its original source as it spreads, losing key bits of information \citep{ribeiroMessageDistortionInformation2019,tanLostPropagationUnfolding2016}. When important information or context about research is lost, it has the potential to mislead the public about basic scientific facts and in turn, harm the integrity of research work and policy decisions. Yet empirical analyses examining how information evolves and decays ``in the wild'' often remain limited to single platforms, although in reality, information spreads across many \cite{hillStudyingPopulationsOnline2019,andersonTeensSocialMedia2018}. Importantly, platforms have different sets of users \cite{horvatBirdsFeatherFlock2021} and affordances, and hence variable effects on how information is shared and displayed \cite{malikIdentifyingPlatformEffects2016}, shaping consumption and engagement. For example, Facebook posts have a high text limit (63,206 characters) but are often much shorter as people are unlikely to read long content on social media. Meanwhile, a blog post has neither text limits nor particular norms about content.

To understand how key information is lost or retained, a multi-platform analysis is thus imperative. 
Using a dataset from Altmetric LLC \citep{altmetricsupportOurDataOur2021}, a service tracking mentions of research online, we can examine mentions of research spread across platforms over time. This massive data collection over the past decade has recorded mentions of scientific articles by their unique identifiers, overcoming a major hurdle of consistent and reliable tracking of information across platforms at scale. 

We use this dataset to study information retention in the millions of online mentions of 9,765 of the most popular scientific articles across five categories of platforms: blogs, Facebook, news, Twitter, and Wikipedia. We develop a novel framework and measure for examining information retention. 
In particular, we propose a framework that looks at spikes of attention \cite[aka bursts, see][]{Kleinberg2003,barabasiOriginBurstsHeavy2005,chengCascadesRecur2016} per platform in temporally-ordered sequences. This approach helps us filter out random noise from individual mentions and instead identify meaningful cross-platform, aggregated moments of online attention to a scientific finding.
To measure information retention, we quantify computationally the information retention in an online post in comparison to the abstract of the scientific article. We evaluate the validity of our measure using data labeled by domain experts.
Using this approach, we ask:

\begin{itemize}
    \item 
    \textbf{RQ1. Information retention over time.} Given that any burst in a sequence may refer to the actual source, is information consistently retained over time?
    \item 
    \textbf{RQ2. Information retention across platforms.} As different types of platforms present constraints about text, content, and posting access, how does the information retention differ across platforms?
\end{itemize}

We find that online discussions of science show a strong propensity towards information \textit{loss} in all platforms but social media. However, early platform differences do not affect long-term levels of information retention.
We present three main contributions: we (1) characterize how online discussions of scientific work tend toward low levels of information retention; (2) highlight the role of \textit{multi}-platform discussions of research; (3) and present a burst-based framework for evaluating information across platforms, adaptable to other contexts with ``bursty'' behavior.

\section{Related Work}
\subsection{The online spread of scientific work}
In the past two decades, the internet has become an increasingly important source of news about scientific research \citep{brossardNewMediaLandscapes2013,vaneperenHowScientistsUse2011,hargittaiHowYoungAdults2018,nationalscienceboardScienceTechnologyPublic2020}. 
Understanding how scientific content spreads online is thus a fundamental and critical question. Prior research on the online spread of science has examined how scientists discuss their work \cite{robinson-garciaUnbearableEmptinessTweeting2017,reinhardtHowPeopleAre2009} and what factors are associated with better scholarly communication \citep{milkmanScienceSharingSharing2014}. More recently, \citet{zakhlebinDiffusionScientificArticles2020} provided an overview of how scientific findings diffuse across various platforms through sharing by users and organizations.

Work looking at the diffusion of online content more broadly has focused on the flow and virality of content on new media platforms \cite{adarTrackingInformationEpidemics2005,adamicInformationEvolutionSocial2016,lakkarajuWhatNameUnderstanding2013,lermanInformationContagionEmpirical2010, romeroDifferencesMechanicsInformation2011,bakshyRoleSocialNetworks2012}, noting the bursty 
behavior of online content \citep{chengCanCascadesBe2014,chengCascadesRecur2016}. This work has also highlighted how information mutates as it spreads, as seen in clickbait and fake news \cite{chenMisleadingOnlineContent2015,vosoughiSpreadTrueFalse2018,lazerScienceFakeNews2018,guptaFakingSandyCharacterizing2013}.
In the case of the online spread of science, prior work investigates what features of scientific articles are linked with more news coverage and sharing \citep{milkmanScienceSharingSharing2014,maclaughlinPredictingNewsCoverage2018}. Importantly, facts often get distorted alongside the loss of key information and context in favor of attracting attention \citep[see for example][]{burnettHowMediaWarp2017}. Thus, understanding information retention in online discussions of science is a critical step for improving science communication on the web. 

In an experiment examining information cascades about medical research outputs, \citet{ribeiroMessageDistortionInformation2019} showed how information distorts over the hops of an information cascade, resulting in content that resembles misinformation even without intent to misinform. 
Existing \textit{empirical} analyses of how information changes as it spreads are based on single platforms and frequently focus on cascades, where one point of a cascade shapes subsequent ones \citep{easleyInformationCascades2010}. Yet, information spreads variably across many platforms and likewise, through various streams that may or may not cascade. To address this gap, we examine information retention across multiple platforms and rather than focusing on cascades, look at information retention in trajectories of bursts of attention as an important step in understanding how discussions of science evolve.


\subsection{Examining multiple platforms}
Two factors complicate how information changes as it spreads online. First, information does not stay within a platform but is free to flow across a connected ecology of platforms. As such, a growing body of work argues for cross- and multi-platform analyses which can produce work that goes beyond particular use contexts to develop higher-level understandings of ecologies of information flow \citep{hillStudyingPopulationsOnline2019,phadkeManyFacedHate2020,zakhlebinDiffusionScientificArticles2020,horvatBirdsFeatherFlock2021}.

Second, the platforms through which information travels have particular designs which may affect the fidelity of the information. In other words, platforms and their design constraints matter \citep{malikIdentifyingPlatformEffects2016,gligoricHowConstraintsAffect2018}. For example, \citet{malikIdentifyingPlatformEffects2016} use temporal data from Facebook and Netflix to give proof-of-concept to the notion of platform effects, demonstrating how use differs on both platforms before and after a significant design change that alters user exposure to information. Prior work also notes how content evolves as it travels across different platforms \citep{leskovecMemetrackingDynamicsNews2009,tanLostPropagationUnfolding2016,phadkeManyFacedHate2020}. In a study on information propagation, \citet{tanLostPropagationUnfolding2016} look at the diffusion of information about press releases on Facebook to describe how information tends to diverge from the source material is it spreads through "layers" of mediums: news articles covering a press release, social media shares, and comments on those shares \citep{tanLostPropagationUnfolding2016}.
\citet{zakhlebinDiffusionScientificArticles2020} investigate the dynamics of cross-platform information diffusion specifically for scientific articles, highlighting how platforms matter in structural virality.


Although information diffuses across platforms and its distortion remains a major point of concern, there is little understanding of the dynamics of how information changes as it propagates across such a multi-platform information landscape. To address this gap, we propose a novel measure and framework to examine multi-platform information retention in the online discussion of scientific articles.

\section{Research design}
\subsection{Data}
\begin{table}[t]
    \centering
    \begin{tabular}{r|r|r|c}
         \textbf{Medium} & \textbf{Tracked posts} & \textbf{Text collected} & \textbf{\%} \\
         \hline
         Blogs & 135,494 & 98,151 & 73\% \\
         Facebook & 130,502 & 94,467 & 72\% \\
         News & 258,367 & 142,606 &  55\% \\ 
         Twitter & 4,426,264 & 3,923,874 & 89\% \\
         Wikipedia & 5,976 & 5,976 & 100\% \\
         \hline
         \textbf{Total} & \textbf{4,956,603} & \textbf{4,264,894} & \textbf{72\%}\\
    \end{tabular}
    \caption{Text we collected for the public posts that mention the selected 9,765 scientific articles according to Altmetric LLC (as of Oct. 2020).}
    \label{tab:collected_data_nums}
\end{table}

We use data from Altmetric LLC, which tracks the multi-platform spread of scientific articles and is the most comprehensive data of this kind for the past decade. ``Platforms'' tracked by the Altmetric LLC dataset include Facebook, Twitter, Google+, Reddit, YouTube, the Stack Overflow network, and Wikipedia in addition to umbrella categories of ``news'' and ``blog'' sites \cite{altmetricsupportOurDataOur2021}. The dataset, starting in 2011, tracks mentions of scientific articles by their respective document identifiers (DOIs), overcoming the challenge of consistent tracking, and has the advantages of covering scientific articles from a wide range of domains and being highly granular.

Given our analytical approach, which relies on examining bursts of attention, studying scientific articles with only a few mentions does not yield meaningful data for investigating information retention at the level of burst sequences. Using both the 2018 and 2019 data dumps\footnote{A dump should subsume previous ones, but counts can decrease when mentions are deleted. We combined the dumps to improve the consistency and reliability of data.} provided by Altmetric LLC, we identified scientific articles that were among the 25,000 most mentioned online in both years. We then selected articles for which we could retrieve abstracts (from Web of Science\footnote{clarivate.com/webofsciencegroup/solutions/web-of-science/} and Microsoft Academic Graph\footnote{microsoft.com/en-us/research/project/microsoft-academic-graph/}), had a minimum abstract length of 500 characters, were in English, and were published in or after 2016 (as Altmetric LLC's method of tracking DOIs changed in 2016). This resulted in 9,765 scientific articles, each of which have at least 200 tracked mentions.

While platforms such as Reddit or YouTube are interesting sites of study, our exploratory investigations showed that---in addition to being rather sparse for a large-scale evaluation---content on these platforms primarily contained links (e.g. to the scientific article) without much additional context, or text could not be obtained (e.g. site was defunct or transcript unobtainable). As our analyses are based on the text of mentions, we focused on platforms for which we could obtain text beyond links, i.e., Facebook, Twitter, news sites, blogs, and Wikipedia.

We collected the relevant online mentions via the Altmetric API in October 2020. We consider public mentions by users of various digital media, both individuals and organizations. Because the data only tracks direct DOI mentions of scientific articles, this is a conservative account of the spread of science information online. Additionally, replies and responses to such mentions are not captured in the dataset; we discuss this point in the Limitations section.
The Altmetric API provided the full text of shorter length mentions. To obtain the full text of longer mentions for each platform, we wrote a series of scripts leveraging APIs and Python libraries to collect public mentions from the blog (posts), news (articles), Facebook (posts), Twitter (tweets), and Wikipedia (article revisions) categories. 
In specific, we used the Twitter and MediaWiki APIs to collect texts for Twitter and Wikipedia, and wrote custom scripts to collect text from blogs, news, and Facebook. For blogs and news, we utilized the news API script\footnote{newsapi.org/docs/client-libraries/python}. For Facebook, about 55\% of posts were short enough such that the Altmetric LLC dump stored that text in full. For the rest, we wrote a custom script to only collect publicly-available posts. 

We removed irrelevant content such as ads in news, meta information, and URLs. We also de-duplicated data to ensure that the same mention instance is not repeated in the dataset. However, we do not de-duplicate mentions with the same text from \textit{different} sources; similarly, we keep retweets. We do so to reflect in our data the fact that content, if disseminated by multiple sources/users, has been seen more widely. Weighing this content more heavily due to its repetition more accurately reflects how a scientific article is being discussed overall. For example, a text spreading via multiple news sources may have a larger impact on public perception than if only one source had shared it.

Overall, we identified nearly 5 million mentions between January 1, 2016 and December 31, 2019 for our 9,765 articles and were able to retrieve text for roughly 72\% of them. Table \ref{tab:collected_data_nums} shows the percentage of mentions for which we collected text in each of the five platforms.


\subsection{Developing a computational measure of information retention}
\begin{table*}[ht]
    \centering
    \begin{tabular}{r|l}
        \textbf{Numbers} & ``September 2012'', ``34 randomised controlled trials'', ``95\%CI = 0.03;0.22)''\\ \hline
         \textbf{Descriptive nouns} & ``exercise'', ``cancer type'', ``adult patients'', ``different demographic and clinical characteristics''\\ \hline
         \textbf{Named entities} & ``QoL'', ``PubMed''\\ \hline
         \textbf{Research design and jargon} & ``control group'', ``treatment'', ``'post-intervention outcome values''' \\
    \end{tabular}
    \caption{Examples of keyphrases extracted from a paper on the effects of exercise on cancer patients.}
    \label{tab:keyphrase_example}
\end{table*}

Whether something has retained key information is nebulous, subject to human judgement. Thus, any computational measure of information retention is imperfect. However, \citet{ribeiroMessageDistortionInformation2019} found that keyphrase survival correlated with evaluations of fact survival. Given this, we devised a simple keyphrase-based computational measure, intended to provide a reasonable evaluation of information retention at scale but remaining scrutable enough to understand the measure's limitations.

We used the TextRank algorithm implemented in the pytextrank library to extract keyphrases from research abstracts as it is one of the top text summarization and keyword techniques, demonstrated on research abstracts \citep{mihalceaTextRankBringingOrder2004}.
The TextRank algorithm creates a lemma graph of noun chunks and named entities (which constitute the keyphrases) based on sentences and part-of-speech tags. It uses this graph to assign each keyphrase a rank value measuring its importance via PageRank (a type of eigenvector centrality). 
Using these ranks, our measure takes an abstract's keyphrases and searches for them in the text of a mention to give an information retention ``score'':
$$\frac{sum(\text{ranks of abstract keyphrases found in post text})}{sum(\text{ranks of all abstract keyphrases})}$$

The rank values in this formula for a given research paper come from a single network, which makes comparisons between score values for the same paper meaningful. A perfect score of 1 means that all keyphrases in the abstract were found in the text of the mention (perfect information retention); a score of 0 means that no phrases were found in the text (complete information loss). 

To better understand the keyphrases in our method, we qualitatively coded phrases from a random sample of 20 abstracts to develop a heuristic of keyphrase types: (1) numbers; (2) descriptive noun phrases; (3) named entities and abbreviations; (4) research design and jargon. Table \ref{tab:keyphrase_example} gives examples of keyphrases from a randomly chosen article from our data set\footnote{This is a meta-analysis article on the effects of exercise on cancer patients, DOI: 10.1016/j.ctrv.2016.11.010}. We found that the online mentions of this example article typically contained ``numbers'' or ``descriptive noun phrases''. Further examples can be found in the Supplementary Materials.

Note that while our measure is based on prior work highlighting the relationship between keyphrases and information retention \cite{ribeiroMessageDistortionInformation2019}, it is not perfect. For example, content might use synonyms rather than the exact phrases used in the abstract. While key scientific terms are unlikely to have many synonyms, in some cases our measure could underestimate information retention. Content could also suffer from information loss while using phrases out of context or omitting other details. Thus, we conducted a survey among domain experts to ensure the validity of the measure for our particular application. 

\subsubsection{Measure validity}
We conducted an IRB-approved survey to compare our measure against human assessments of information retention. Because this study focuses on information from scientific research, we asked academics to respond to our survey in order to improve confidence in the computational measure.

We recruited academics through listservs and department contacts. We asked participants for their broad research area (biomedical, social, and physics \& engineering) and presented to them a relevant set of 5 research abstracts. Per abstract, participants were given 1 to 5 pairs of online mentions of the research (each pair being specific to a platform) and asked to select which mention they believed had more information loss, defined as ``losing important information and/or key facts from the original reference text'' (the abstract). Participants also had the option to answer that ``I believe the two texts have the same amount of information loss''.

Abstracts were selected manually by the research team from the 9,675 articles with the following criteria: the abstract was between 500 and 2000 characters in length, so that participants could reasonably read it; the set of 5 articles covered a diverse range of topics within the broad research area. 
Per article, we selected 1-5 platforms in our dataset that the article was mentioned on, with the goal of having multiple survey items per platform. For each platform chosen for a scientific article, we selected two online mentions after running the computational measure over all of them: one from the highest resulting scores, and one from the lowest. 

\begin{table}[t]
    \centering
    \begin{tabular}{@{}l|r|r@{}}
     & \textbf{Match} & \textbf{High-agreement cases} \\ \hline
    \textbf{Overall} & \textbf{62\%} & \textbf{90\%} \\ \hline
    {Blogs} & 58\% & 85\%\\ 
    {Facebook} & 72\% & 85\% \\ 
    {News} & 55\% & 90\% \\ 
    {Twitter} & 63\% & 100\% \\ 
    {Wikipedia} & 57\% & 93\% 
    \end{tabular}
    \caption{Percentage of survey responses matching with the computational measure's assessment, by platform.}
    \label{tab:survey_items_platforms}
\end{table}


Our survey yielded 30 responses from 19 PhD students, 7 post-docs, 3 faculty (assistant professors), and 1 researcher who wrote ``molecular biologist" as their position. The survey contained 72 survey items (mention pairs for assessment), each evaluated by five participants. We focused on two outcomes: how the measure performed in the survey overall and how the measure performed in the survey when expert agreement was high (i.e., when human assessment was more certain).
Overall, 62\% of responses matched our computational measure's assessment, compared to a random baseline of 33\%. 60\% of survey items were ``high-agreement'' cases, where at least 4 of 5 people had the same response. For these cases, we found that 90\% of the responses matched our computational measure, with a bootstrapped 95\% confidence interval of (86\%, 94\%). Table \ref{tab:survey_items_platforms} shows these numbers disaggregated by platform. 
In sum, the survey validates our computational measure as it performs much better than the baseline overall, and particularly well when expert agreement is high. It also notes how information loss can be difficult for humans to assess. We discuss the implications of our measure in the Limitations section.

A stronger validation of our measure might present random mentions to respondents, instead of those with highest versus lowest information retention scores. However, we made this simplification given the limited availability of experts and the fact that our goal was primarily to assess if the measure carried a strong enough signal compared to human judgement. To counter the limited scale of expert validation, we further tested the robustness of our findings using an alternative keyphrase extraction method.

\subsubsection{Robustness check with alternative method} We also ran our analyses with a measure based on another well-known keyphrase extraction algorithm, {\emph{Rapid Automatic Keyword Extraction (RAKE)}} \cite{roseAutomaticKeywordExtraction2010}. We found that the most notable patterns obtained with our TextRank-based measure are replicated with RAKE. Figures \ref{fig:sequences2_9} and \ref{fig:burst_medianscore_platform} include results obtained with both TextRank and RAKE; all replicated analyses are available in the Supplementary Materials.

\subsection{A burst-based framework}

\begin{figure}[t]
    \centering
    \includegraphics[width=0.95\columnwidth]{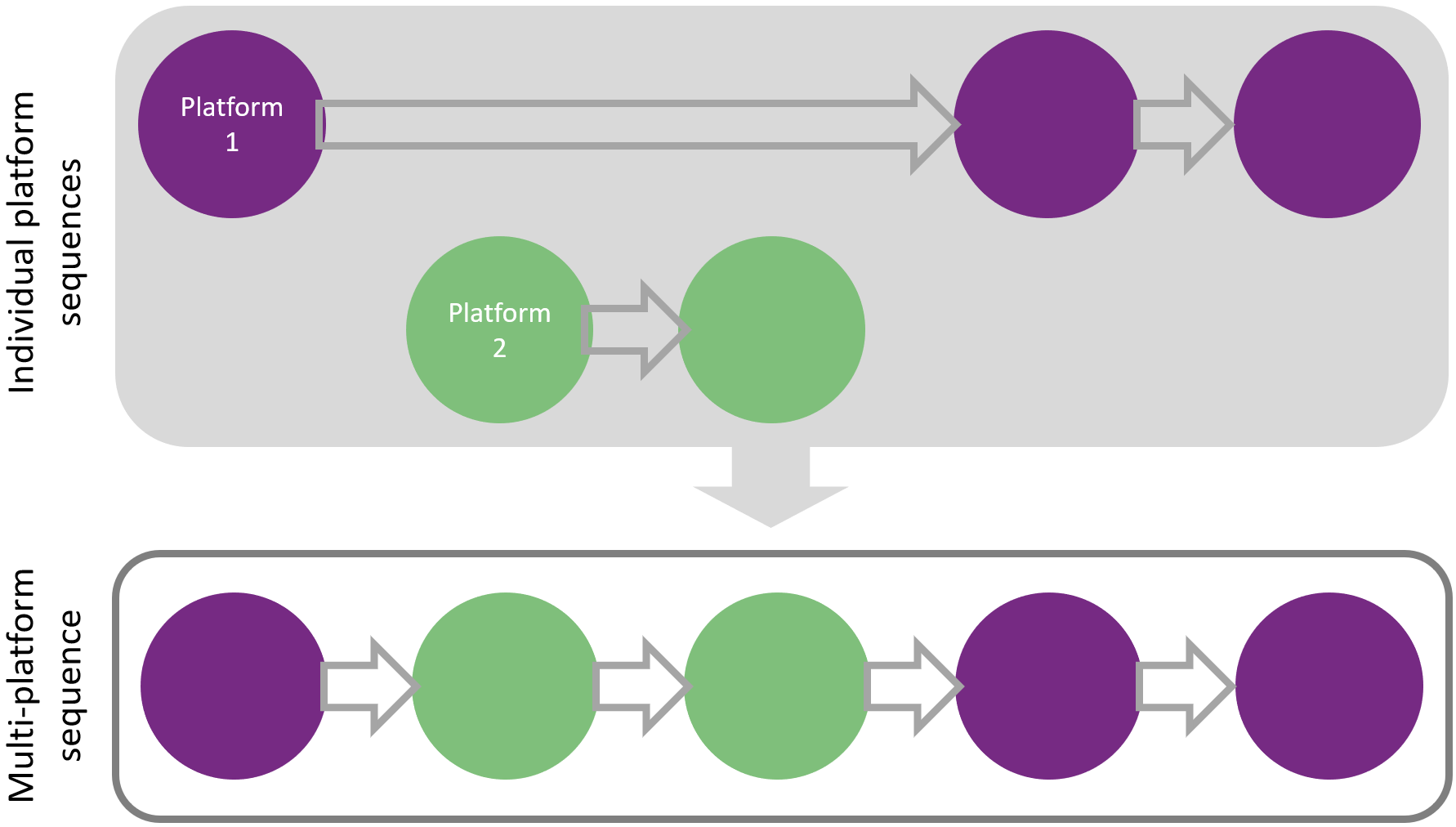}
    \caption{From burst sequences on individual platforms to a multi-platform sequence of bursts per article.}
    \label{fig:burst_example}
\end{figure}

Taking inspiration from the ``hop''-based analysis of information cascades in \citet{ribeiroMessageDistortionInformation2019} and the bursty nature of online content \citep{zakhlebinDiffusionScientificArticles2020,gilbertWidespreadUnderprovisionReddit2013,chengCascadesRecur2016}, we present a burst-based framework for evaluating information retention across platforms ``in the wild''. 
Rather than evaluating individual mentions of a scientific article, we focus on when an article receives a concentrated peak in attention, or a ``burst'' of online mentions. Bursts present a strategy to bin and aggregate data, providing a principled way to extract meaningful chunks as units of analysis.
In particular, they present essential moments of online discussion in time about a paper, reduce random noise, and facilitate computational analysis. Creating bursts for each platform also results in comparable units across platforms rather than individual posts of arbitrary form. We construct burst sequences for each platform per research article, following the parameters outlined in \citet{chengCascadesRecur2016}. This approach identifies a burst on a given day when the daily number of mentions of the monitored content (here, a research article) meets a set minimum count and is unusually high relative to the number of mentions on surrounding days.

\begin{table}[t]
    \centering
    \begin{tabular}{r|l|l|l|l|l}
    & \textbf{Blogs} & \textbf{Faceb.} & \textbf{News} & \textbf{Twitter} & \textbf{Wiki.} \\
    \hline
    Min. \# & 7 & 8 & 8 & 9 & 7 \\
    \end{tabular}
    \caption{Weighting the daily number condition. Minimum number of mentions required on different platforms.}
    \label{tab:condition_1_weights}
\end{table}

As the distribution of mentions of research outputs across platforms is highly skewed---with Twitter dominating in the dataset---we make a light alteration to the conditions specified in prior work \citep{chengCascadesRecur2016} by creating different weights for platforms, such that bursts on platforms with higher barriers to posting---and thus, fewer posts overall---require fewer posts at a given time to be considered a burst. 
The new thresholds for the daily minimum mention count condition (Table \ref{tab:condition_1_weights}) are calculated by taking the logarithm of the count of posts per platform and dividing it by the logarithm of the count of posts across all platforms. We multiply this fraction by the original threshold (10) and round up. The result is the minimum number of daily posts required for the platform to meet the daily minimum mention condition.


\begin{table*}[t]
    \centering
    \begin{tabular}{@{}r|r r r r r @{}}
    & \textbf{Blogs} & \textbf{Facebook} & \textbf{News} & \textbf{Twitter} & \textbf{Wikipedia} \\
    \hline
    \textbf{\# Bursts} & 2,281 & 1,161 & 4,184 & 29,380 & 26 \\
    \textbf{\% \textit{all} mentions being in bursts} & 20\% & 12\% & 51\% & 35\% & 11\% \\
    \textbf{\# mentions in bursts (\% text collected)} & 26,652 (69\%) & 16,172 (89\%) & 130,691 (42\%) & 1,566,438 (77\%) & 631 (100\%) \\
    \textbf{\# English mentions in bursts (\%)} & 22,977 (68\%) & 12,479 (91\%) & 110,059 (44\%) & 1,200,906 (100\%) & 631 (100\%)\\
    \end{tabular}
    \caption{Basic statistics of burst counts on each platform.}
    \label{tab:burst_post_counts}
\end{table*}


We generated 37,032 bursts encompassing 1,347,052 English-language posts. The number of bursts per platform and the number of mentions in those bursts as well as the percentage of text data we were able to collect for each platform can be seen in Table \ref{tab:burst_post_counts}. 
Of our 9,765 research papers, 9,616 have at least 1 burst. The median number of bursts per research article is 3 and the mean is 3.85; this number is also highly skewed across research papers.

After calculating bursts per platform, we created a single ``master'' sequence that contains all of a research article's bursts in temporal order, visually demonstrated in Figure \ref{fig:burst_example}. These sequences capture attention over time. Note that in the lack of information transfer between adjacent bursts, burst sequences could be but are not necessarily information cascades (i.e., such that a burst shapes subsequent bursts). This allows us to look across platforms that do not have direct indicators of spread (e.g. Twitter’s Retweets) and is more appropriate as we are interested in tracking attention to research articles in the form of online mentions.

\begin{figure}[t]
    \centering
    \includegraphics[width=0.85\columnwidth]{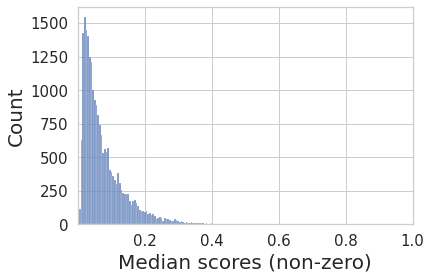}
    \caption{The distribution of burst-level information retention scores that are non-zero. Scores that are '0' make up 32.5\% of burst-level scores.}
    \label{fig:score_distribution}
\end{figure}

We ran our measure on each mention text and took the median of the scores in a burst to produce the burst-level information retention score. This score represents the \textit{median} information retention \textit{per mention} in a burst. An alternative burst-level score construction might be taking the text of all mentions in a burst as one and then computing our measure. Although the two versions were highly correlated, the latter could inflate the burst-level score by including keywords that appear in just one of many mentions of a burst. We found that the former version of the burst-level score based on medians produces lower scores, but is a more faithful representation of information retention of mentions and the values need only be examined relative to one another.
The selected burst-level information retention scores were skewed to the right, as seen in Figure \ref{fig:score_distribution}. 32.5\% of the scores were zero and the highest score was 0.89.



\section{Results}

\begin{table*}[t]
    \centering
    \begin{tabular}{@{}l |r | r r r r r @{}}
    & & Blogs & Faceb. & News & Twitt. & Wiki. \\
    \hline
    \multirow{2}{7em}{All bursts, size} & Mean & 11.7 & 13.9 & 31.2 & 53.3 & 24.3 \\
    & Median & 10 & 11 & 23 & 23 & 11.5 \\
    \hline
    \hline
    \multirow{5}{7em}{First} & \# Bursts & 1318 & 463 & 2214 & 8239 & 4\\
    \cline{2-7} 
    & Score & 0.13 & 0.04 & 0.08 & 0.03 & 0.02 \\ 
    \cline{2-7} 
    & \% All bursts & 58\% & 40\% & 53\% & 28\% & 15\% \\
    \cline{2-7} 
    & \% First bursts & 11\% & 4\% & 18\% & 67\% & 0\% \\
    & \textit{Expected \%} & 6\% & 3\% & 11\% & 79\% & 0\% \\
    \hline
    \hline
    \multirow{3}{7em}{Co-occur} & \# Bursts & 1,709 & 653 & 1,784 & 1,933 & 4\\
    & \% All bursts & 75\% & 56\% & 43\% & 7\% & 15\% \\
    \cline{2-7} 
    & Score & 0.12 & 0.04 & 0.08 & 0.02 & 0.05 \\ 
    \end{tabular}
    \caption{Burst characteristics: (1) sizes of all bursts, (2) the number of bursts first in a sequence across platforms, along with the median scores and what percentage of all and first bursts this constitutes, and (3) the number of co-occurring bursts.}
    \label{tab:first_co_occurring}
\end{table*}

\begin{figure}
    \centering
    \includegraphics[width=0.7\columnwidth]{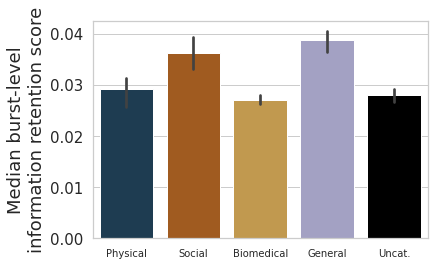}
    \caption{The median of burst-level information retention scores across main discipline areas.}
    \label{fig:burst_domains}
\end{figure}

We first contextualize our results by noting basic descriptives of bursts. 
As shown in the top part of Table \ref{tab:first_co_occurring}, burst size (i.e. the number of mentions in a burst) varies substantially across platforms and covers a wide range. On each platform, burst size is highly skewed to the right. Figure \ref{fig:burst_domains} shows the median burst-level information retention score across broad disciplines categorizing articles, relying on the SCOPUS labels that the Altmetric LLC data includes. In our set of 9,765 articles, 56\% of articles were in Biomedical Sciences, 6\% in Social Sciences, 5\% in Physical Sciences, 14\% in General, and 28.5\% in Uncategorized. Physical Science, Biomedical Science, and Uncategorized articles had similar median scores (0.029, 0.027, 0.028 respectively), lower than General and Social Science articles (0.039, 0.036 respectively).

\subsection{Information retention over time}

\begin{figure*}[ht]
    \centering
    \includegraphics[width=0.85\textwidth]{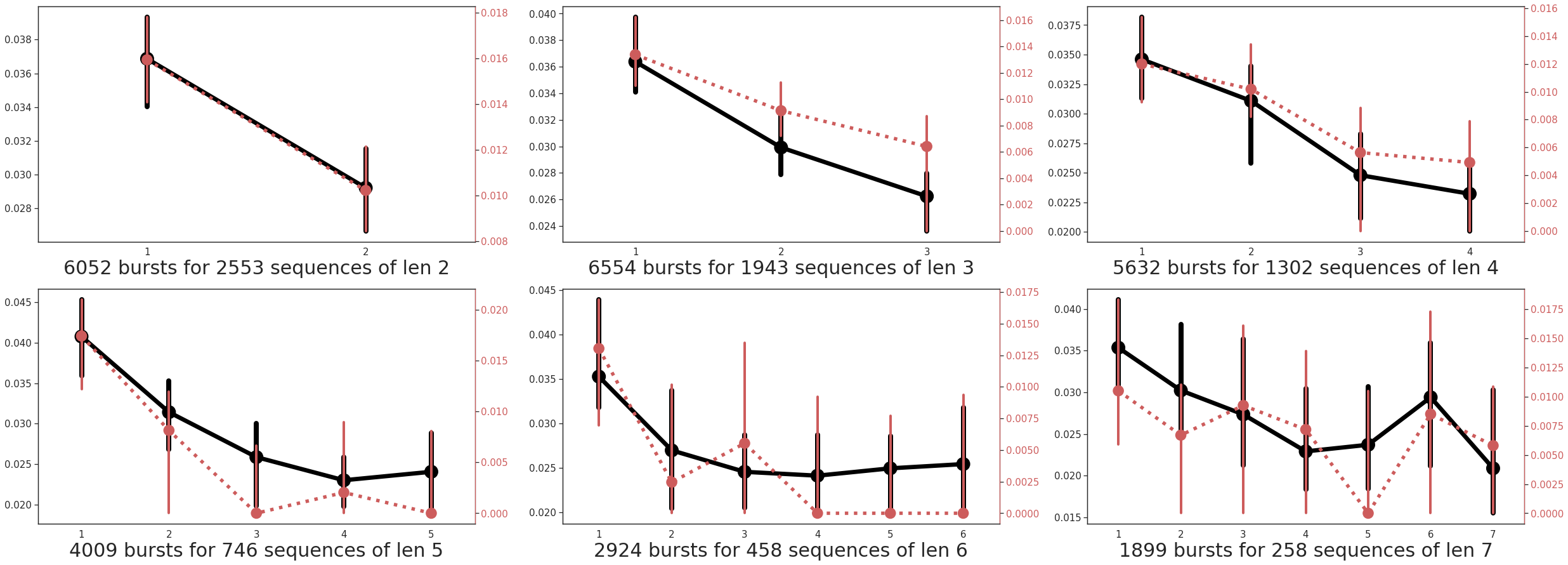}
    \caption{Trajectories of information retention scores for burst sequences of a certain length, with 95\% CI around the median for that point in the sequence. The dotted red line shows the trajectory using the RAKE-based score as a robustness check.}
    \label{fig:sequences2_9}
\end{figure*}

To examine information retention over time, we look at burst sequences with a minimum length of 2. Note that co-occurring bursts occupy the same sequential point, e.g., a sequence with just two co-occurring bursts has a length of one. Burst sequences range in length from 1 through 82, with a median length of 3; 8,193 research papers have at least two bursts. However, the distribution of bursts of a certain length is right-skewed, and we narrow our analyses to sequence lengths such that the final burst in the sequence has at least 200 cases, creating an upper limit of length 7.

In Figure \ref{fig:sequences2_9}, we show the trajectory in information retention score for sequences of lengths 2 through 7, specifically calculating the median score of bursts at each position in the sequence, for all sequences of that length. The figure shows that for all sequence lengths, retention scores usually continuously drop or stay the same as the prior burst in the sequence. While the confidence intervals of adjacent bursts overlap, there is a clear decline in early versus late bursts, with a steeper drop in earlier positions of sequences before leveling off. 
This trend is unexpected. Remember that bursts in a sequence are simply temporally ordered: any given burst may not be a direct offshoot from the burst coming before it, but can refer to the original source text, the abstract. Interestingly, despite the fact that references to the abstract are possible at any burst, bursts never see jumps in their information retention trajectory that would rival the first burst. Though the data cannot tell us whether there is information transfer as in an information cascade, the continued decline in information retention across sequences suggests that later bursts do not refer back to the source material.

The first burst may hold a special role in information retention trajectories, as it consistently has the highest scores in the sequences. First bursts are also of interest because they indicate the platform in which concentrated attention to an article first appeared online, and thus, may set the tone for the public discussion of the paper. The number of first bursts across platforms can be seen in the middle section of Table \ref{tab:first_co_occurring}. In terms of the platform on which these first bursts are distributed, Twitter again has the most bursts (making up 67\% of all of the first bursts) and is followed by news (18\%), blogs (11\%), Facebook (4\%), and then Wikipedia (0\%). Based on the number of posts per platform, this distribution is relatively unsurprising. However, looking at the distribution of bursts across platforms, we might actually expect Twitter to be higher in what percent of the first bursts it makes up (79\%) and news and blogs to be lower (11\% and 6\%, respectively). In other words, news and blogs platforms make up more of the first bursts than expected. Moreover, we see that 58\% of all bursts on blogs and 53\% of all bursts on news are first bursts. This means that for blogs and news, a majority of the bursts come early on in the online discussions of science.

\subsection{Information retention across platforms}

\begin{figure}[t]
    \centering
    \includegraphics[width=0.9\columnwidth]{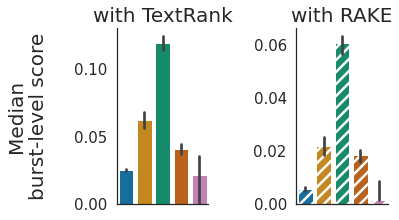}
    \caption{The median burst-level scores per platform.}
    \label{fig:burst_medianscore_platform}
\end{figure}

Figure \ref{fig:burst_medianscore_platform} shows the median information retention score per burst in each platform. This median score is highest in blogs (0.12) followed by news (0.06) and Facebook (0.04), and lowest in Twitter (0.02) and Wikipedia (0.02). Given that news and blog mentions are longer and therefore more likely to contain more keywords, their higher retention scores are not surprising. However, these tendencies in information retention \emph{cannot} simply be attributed to differences in length across platforms. For example, although blogs have twice the median score as news, the median length of news mentions is 670 words versus 656 words on blogs. While Wikipedia and Twitter have equally low information retention scores, Wikipedia mentions have a median length of 126 words while Twitter posts typically contain 17 words.

\begin{figure*}[t]
    \centering
    \includegraphics[width=0.85\textwidth]{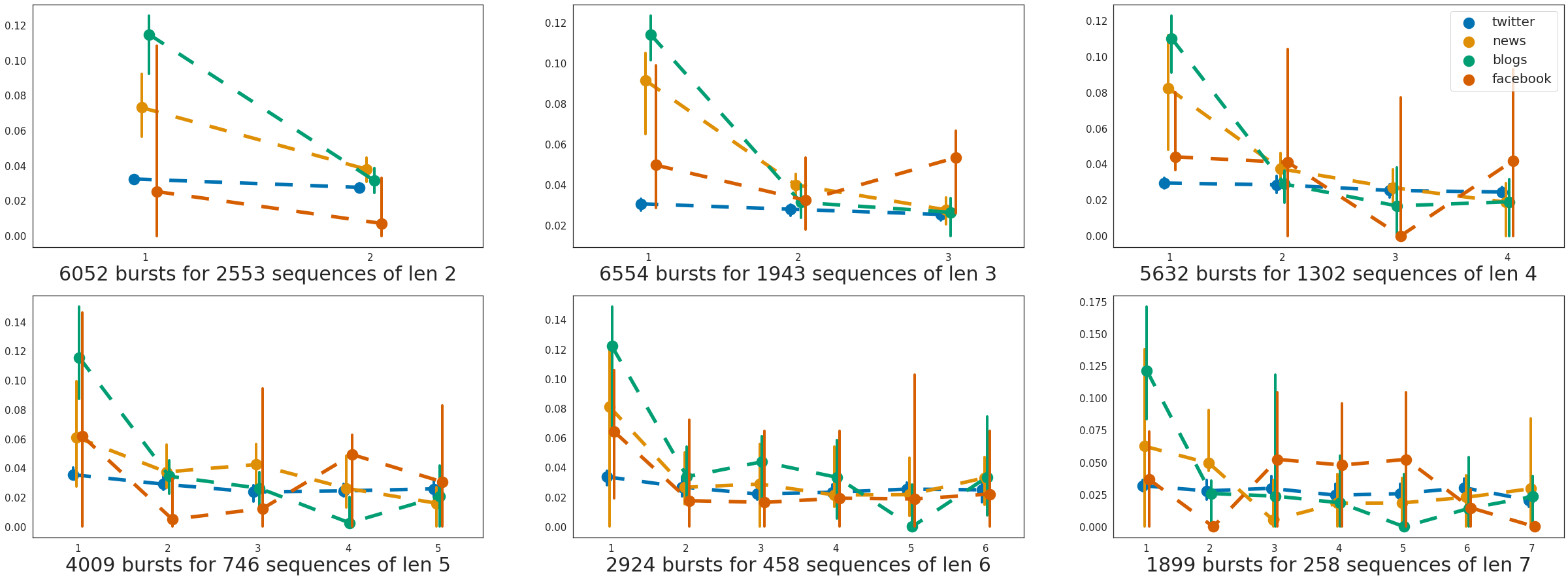}
    \caption{Median trajectories of information retention scores for burst sequences starting in a given platform, using the TextRank-based score. The error bars show the 95\% confidence interval around each median value for that point in the sequence.}
    \label{fig:sequences2_9_first}
\end{figure*}

Given these platform differences, one might expect that sequences starting on different platforms will follow different trajectories. Yet we find that this is not necessarily the case. In Figure \ref{fig:sequences2_9_first}, we recreate Figure \ref{fig:sequences2_9} stratified by the platform on which the \textit{first} burst in the sequence occurred (using just the TextRank-based score). For sequences of all lengths, first bursts from different platforms have different median information retention scores, with blogs typically exhibiting the highest scores. However, regardless of where the sequence starts, scores drop to near-similar low levels of information retention. The exceptions to this pattern of ``clear drop'' are sequences that start on Facebook and Twitter; in these cases, the first bursts already start off relatively low and remain low throughout the trajectory. 
This suggests that early platform differences do not make a difference in the long-run for information retention. Moreover, the consistently low information retention of sequences starting on Twitter and Facebook suggests that later bursts more often consist of social media bursts than news or blog bursts.

\begin{figure}[t]
    \centering
    \includegraphics[width=0.8\columnwidth]{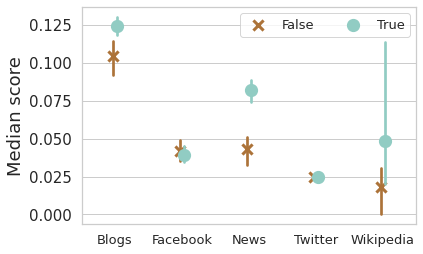}
    \caption{Median scores of bursts that are co-occurring (true) and bursts that are not co-occurring (false), across platforms.}
    \label{fig:co_occurring}
\end{figure}

How do platforms matter, then, along burst sequences, beyond differences in median scores? 
To this end, we examine co-occurring bursts, or bursts on the same article occurring on the same day. Co-occurring bursts suggest that an article received widespread attention that is apparent across multiple platforms. 6,083 bursts are co-occurring, representing roughly 16\% of all bursts, with a breakdown per platform in the bottom part of Table \ref{tab:first_co_occurring}. The distribution of the number of co-occurring bursts across platforms in general follows the pattern for first bursts and number of bursts more broadly. 
Comparing the median scores of bursts that co-occur to those that do not (see Figure \ref{fig:co_occurring}) reveals that co-occurring bursts tend to have significantly higher information retention scores across platforms, except in Facebook and Twitter, where the scores are nearly the same. In other words, bursts of attention involving multiple, non-social media platforms tend to have higher information retention.

\begin{figure}[t]
    \centering
    \includegraphics[width=0.8\columnwidth]{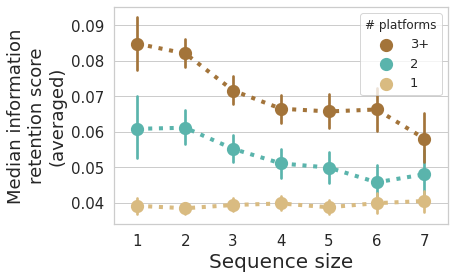}
    \caption{Median scores for sequence lengths 2-7, stratified by how many platforms are in the sequence. Co-occurring bursts may result in more platforms than bursts in sequences.}
    \label{fig:heterogeneity}
\end{figure}

We investigate this further in Figure \ref{fig:heterogeneity}, which shows the median information retention score for sequences of various lengths, stratified by how many unique platforms are in the sequence. Importantly, for every sequence length, sequences with more platforms generally have higher information retention scores. Earlier we noted that news and blog mentions are often concentrated in first bursts. In this sense, since news and blog mentions have a higher median information retention, the decline in score is partly attributable to the fact that bursts later in the sequence are increasingly on platforms with stronger length constraints such as Twitter or lower median scores such as Wikipedia. However, Figure \ref{fig:heterogeneity} suggests that the platforms also matter in a broader sense, such that online discussions held across multiple platforms retain higher-fidelity information.

\section{Discussion}
With heightened discussions around scientific misinformation \cite{scheufele2019science,westMisinformationScience2021}, fundamental knowledge characterizing how research is shared online---where some audiences are as likely to engage with science as with lighter topics \cite{hargittaiHowYoungAdults2018}---is imperative. Research on online science dissemination is quickly emerging \cite{zakhlebinDiffusionScientificArticles2020} 
but so far lacks an understanding of how this information deteriorates as it is shared. To fill this gap, we studied information retention about scientific research articles \textit{over time} and \textit{across platforms} via a novel computational measure applied to large-scale observational data on online attention to nearly 10,000 research articles. As such, our work responds directly to a recent call in the social sciences to develop new approaches that enable creating meaning from Web trace data \citep{lazerMeaningfulMeasuresHuman2021}. Conducting this investigation for online mentions on platforms as varied as Twitter, Facebook, blogs, news sites, and Wikipedia is an important first, and enriches decades of literature on information diffusion \citep{adarTrackingInformationEpidemics2005,lermanInformationContagionEmpirical2010,keeganHotWikiStructures2013,gilbertWidespreadUnderprovisionReddit2013,Goel2015,chengCanCascadesBe2014,chengCascadesRecur2016} with the important albeit less studied aspect of cross- and multi-platform effects \citep[for exceptions, see][]{leskovecMemetrackingDynamicsNews2009,tanLostPropagationUnfolding2016,zakhlebinDiffusionScientificArticles2020}.

\subsubsection{Propensity towards information loss}
Our work shows a strong propensity towards information loss, regardless of how many platforms are in the sequence. This is especially true for sequences starting on news or blogs sites; sequences starting on Facebook and Twitter generally start and stay low in information retention. This consistent trajectory towards low information retention raises concerns given that information loss can often lead to miscommunication of science online. While the identification of the mechanisms explaining these patterns are beyond the scope of this paper, we propose possible explanations as important future directions.

One mechanism driving decreasing retention might be that a research finding naturally becomes less relevant over time; later bursts may only mention it in brief passing, in connection to a separate topic that is the main focus of the mention. Future work examining the centrality of a research finding per mention can shed light on the relationship between decreasing information retention and relevance.

Another potential explanation is a platform-based one: while Twitter (which has the lowest median information retention of the platforms) makes up the majority of bursts at all steps of a sequence, later bursts are typically more heavily dominated by Twitter than earlier ones, which instead tend to also include blogs, news, and Facebook bursts. Similarly, we note that bursts that start on Twitter or Facebook already start with relatively low information retention. This points to the importance of platforms such as blogs and news in improving high-fidelity information across bursts. Moreover, it is possible that press releases describing research findings increase information retention in earlier bursts, particularly as part of news or blog content. However, a manual inspection of 
100 random news and blog mentions in first bursts yielded 1 official press release. Future work might more systematically characterize content per platform and its relationship to information retention.

\subsubsection{Differences in platforms}
Our findings indeed highlighted differences in information retention across platforms. Blogs performed significantly better than any other platform, with a median information retention score twice the size of the second best-performing platform, news. Some of the patterns we saw were partly explained by the typical length of content in the platforms: platforms with longer mentions intuitively have higher information retention. For example, mentions on news and blogs are longer than posts on Twitter and Facebook and therefore have higher retention scores. However, differences in platforms do not perfectly correlate with content length: blogs, for example, had a median score twice that of news despite the fact that the median length of news mentions is slightly larger than blogs. Moreover, we also saw that bursts being on certain platforms had little effect on the information retention of adjacent bursts and long-term information retention.

However, we also note that the tendency towards more information loss over time was substantially less so for sequences starting in Facebook and Twitter, which were relatively flat over time. A closer look at earlier and later bursts on Twitter did not yield substantive differences in terms of mention length or score. The flatter trajectory of Facebook and Twitter bring support to the explanation that lower information retention later on in sequences are driven by platform differences; that is, that the lower-fidelity nature of social media content may overall drag down information retention over time as social media comes to dominate conversation. Future work should test how content between platforms may shape one another, such as testing how introducing higher fidelity content on one platform might affect the qualities of content on another.

\subsubsection{Encouraging multi-platform discussions}
The differences in platforms, particularly with the flatter trajectories of Facebook and Twitter, emphasize that platforms matter. For example, we found that sequences with more platforms tended to have higher information retention scores, at all sequence sizes. While these patterns reveal no causality, they do highlight that cross-platform online discussions of science may be important for richer, more contextualized discussions of research articles. One possible explanation may be that when multiple platforms are active at the same time, there is a greater diversity of actors, content, media coverage, and information sources for people to interact with, which could improve the collective ability of individuals to parse important information about scientific findings.
Encouraging multi-platform discussions of science may therefore promote more accurate and reliable engagement with scientific information. For sequences that start on social media, for example, encouraging multi-platform discussions in later sequences may instead improve information retention. 
On the other hand, being on multiple platforms is a marker of successful spread; articles mentioned on multiple venues may have been written more appealingly and clearly, making it easier to distill a core message and improving information retention online overall. 
These point to at least two potential strategies for improving information retention: first, encouraging active and recurrent dissemination by researchers across outlets beyond social media; and second, promoting the sharing of research findings on \textit{multiple} platforms \textit{concurrently} to diversify the collective voices engaged in online discussion of the work.

\subsection{Limitations and future work}
An important limitation of our work is that information retention as a construct is difficult to quantify, subject to human sense-making and judgement. However, as our measure is simple, it is also much more scrutable than complex approaches---important when interpreting results given the subjective and potentially biased nature of assessing information retention. Moreover, our measure is able to capture a reasonable estimate of how \textit{key} information is retained which is essential for large-scale analyses as reflected in our validation survey. Our measure performs much better than a random baseline, and does extremely well when human experts also agree in their assessments. Moreover, results were consistent when changing the keyphrase extraction method. Regardless, future work might develop more refined measures for information retention, such as also capturing synonyms of keyphrases.

Further, we focus on research papers that are among the most mentioned online in two years, which is non-representative of ``typical'' papers. Regardless, our sample covers a broad set of research topics, a wide range of mention counts, and an important test set for developing frameworks to understand information retention in diffusion, as they represent, by definition, the papers most likely to circulate widely and have impact. Future work might investigate ``typical'' trajectories in sharing science online \cite[similar to patterns of structural virality in][]{zakhlebinDiffusionScientificArticles2020} or, alternatively, patterns of information retention in other types of ``bursty'' online content \citep{gilbertWidespreadUnderprovisionReddit2013,chengCanCascadesBe2014,chengCascadesRecur2016}.

We also note limitations of the dataset and framework. Our data consisted of direct DOI mentions of research articles, which excludes comments or other related content that may be part of the discussion of an article. This has implications in particular with respect to Twitter ``threads'' used to get around character limits\footnote{In 2020, the Twitter API added a ``conversation ID'' feature that enables researchers to pull all tweets that are in response to a given tweet; future work might extract threads summarizing research findings from these conversations.}. However, in our dataset the median character length of Twitter mentions was 107, far below the character limit of tweets, suggesting that mentions are typically not threads. Indeed, examining a random sample of 100 tweets in our data yielded only 3 that were in threads of any sort. Moreover, our method is consistent over time and across platforms (e.g., we do not collect replies or similar threads of Facebook, news, or blogs content).

Further, we use a simple keyword-based approach over other techniques such as topic modeling, which might capture more context around the diffusion process and help identify the \textit{types} of discussions associated with high information retention. However, topic modeling at scale also poses challenges such as the need for manual inspection to label topics, especially with so many scientific fields. Such methods also tend to be noisy when applied to short documents like social media posts. For this study, we focus---as an initial step---on overall cross-platform information retention independent of the details of the surrounding context in which papers are mentioned. A natural next step for future work building on our contributions includes examining such context and its relationship with information retention.

Finally, we do not make empirical claims to causality. While we identify clear differences across platforms and trends across them, we cannot say, for example, that an increase in the number of platforms involved in a sequence \emph{leads to} higher information retention. In part, this is because our sequences are not necessarily information cascades transferring information from one to another, but simply temporally ordered. However, this study does not set out to disentangle particular effects, such as the ``telephone effect'' \citep{ribeiroMessageDistortionInformation2019}, but instead provides an overview of patterns over time in a multi-platform information landscape. Exciting future research directions include parsing out the mechanisms by which platforms affect information retention. One potential such project includes a systematic analysis of the kinds of keywords found in mentions at various points in a sequence, which may not only articulate particular platform effects but also elucidate why the presence of multiple platforms is linked with higher information fidelity overall.

\section{Ethics Statement}
A serious consideration for the presentation of any quantitative measure is how it may incorporate biases or black-boxes of complex human concepts. As a difficult construct to quantify, information retention is a par excellence example of this, as noted in our discussion of the validation survey. In order to prevent our proposed quantification from becoming misapplied, we selected the most scrutable version of it, such that interpretation and limitations of what the measure suggests are clear. This simplicity also helps undermine unreasonable extrapolations in arguments about the \emph{quality of information retention} of content when applied beyond this work.
Additionally, research utilizing posts from sites containing content from individuals, such as social media and/or blog users, who may not be aware that their content is used for research also encounter important concerns about the integrity of users' privacy. In our work, we consider only public-facing content accessible by any arbitrary individual. In addition, we only use posts that were still available and not deleted by the users at the time of our study. Finally, our results describe user behavior only in aggregate and at a scale that leaves individual posts unidentifiable.

While minimizing potential risks, the expected benefits of our contributions to the understanding of information retention in the diffusion of scientific findings across platforms are substantial and foundational for a better appreciation of intentional and unintentional information distortion online. Our findings not only point to general patterns of information retention that might inform media strategies, such as deliberate dissemination across multiple channels, but also are generative in raising potential mechanisms for improving the fidelity of information in online discussions to be tested and examined for causality in future studies.


\section{Conclusion}

As scientific findings spread on the web, they are discussed across multiple platforms, shaping what information is retained. Accurately communicating science online has critical implications for policymakers, researchers, and the public alike, but the difficulty in multi-platform data collection has made it extremely challenging to unpack how crucial information is retained in a multi-platform information landscape. In this study, we utilized a large-scale observational dataset that leverages unique identifiers of scientific work (DOIs) to track content across different platforms, and examined information retention in bursts of attention to scientific articles over time. Our study offers three main contributions. First, we provide a view of how online discussions of scientific findings lose information ``in the wild'', showing a strong propensity for low information retention. This underscores an important need to devise strategies to mitigate such loss. Second, to this end, we show that scientific articles discussed on more platforms tend to have higher information retention. This suggests that \textit{multi-}platform discussions may help improve information retention and highlights future directions to untangle the mechanisms driving this trend. More broadly, this dynamic also highlights that multi-platform work is critical to understanding how online activity inherently shapes societally-relevant information. Finally, we provide a simple, scrutable measure that can reasonably evaluate information retention at scale and a burst-based framework for applying it to study diffusion in science and beyond. Along with our findings, the measure and framework lay the foundations for further work evaluating the quality and fidelity of information for various types of online content. In a time with ongoing debates about what is factual, understanding how information communicated on the web changes as it spreads over time and across platforms is a pressing societal challenge.

\section{Acknowledgments}
The authors would like to thank students Amanda Hardy and Joshua Jacobs for their invaluable assistance in data work, Hao Peng for generously sharing code, the Community Data Science Collective for earlier feedback, and reviewers for their thoughtful comments that strengthened this manuscript. This work was supported by the NSF (DGE1842165, IIS-2133963, IIS-2133964, and CAREER Award IIS-1943506) and by the Air Force Office of Scientific Research under award number FA9550-19-1-0029.

\bibliography{manuscript}

\end{document}